\documentclass[12pt]{article}

\usepackage{amsmath,amssymb,amsfonts,color,graphicx,cite,color,soul}

\input paperdef

\graphicspath{{figs/}}

\evensidemargin \oddsidemargin
\allowdisplaybreaks

\begin{document}


\thispagestyle{empty}
\setcounter{page}{0}
\def\thefootnote{\fnsymbol{footnote}}

\begin{flushright}
arXiv:1211.4784 [cs.DC] \\
\end{flushright}

\mbox{}\vspace{2em}

\begin{center}

{\large\sc {\bf Phenomenology Tools on Cloud Infrastructures using}}

\vspace*{0.3cm}

{\large\sc {\bf OpenStack}}

\vspace{0.5cm}

{\sc I.~Campos$^{\,1}$%
\footnote{
email: isabel.campos@csic.es
}%
, E.~Fern\'andez-del-Castillo$^{\,1}$%
\footnote{
email: enolfc@ifca.unican.es
}%
, S.~Heinemeyer$^{\,1}$%
\footnote{
email: Sven.Heinemeyer@cern.ch
}%
,\\[.3em] A.~Lopez-Garcia$^{\,1}$%
\footnote{
email: aloga@ifca.unican.es
}%
, F.~v.d.~Pahlen$^{\,1,2}$%
\footnote{
email: pahlen@ifca.unican.es
}%
\footnote{MultiDark Fellow}%
~and G.~Borges$^{\,3}$%
\footnote{
email: goncalo@lip.pt
}%
}

\vspace*{0.8cm}

$^1$ Instituto de F\'isica de Cantabria (CSIC-UC), 
     E--39005 Santander, Spain

\vspace*{0.3cm}

$^2$ Instituto de Biocomputacion y Fisica de Sistemas Complejos - BIFI \\
University of Zaragoza, E--50200, Zaragoza, Spain

\vspace*{0.3cm}

$^3$ Laboratorio de Instrumentacao e Fisica Experimental de Particulas - LIP \\
PT--1000-149  Lisboa, Portugal

\end{center}

\vspace*{0.5cm}

\begin{abstract}

We present a new environment for computations in
particle physics phenomenology employing recent developments in
cloud computing. 
On this environment users can create and manage ``virtual'' machines on
which the phenomenology codes/tools can be deployed easily in an
automated way. We analyze the performance of this environment
based on ``virtual'' machines versus the utilization of 
physical hardware. 
In this way we provide a qualitative result for
the influence of the host operating system on the performance of a
representative set of applications for phenomenology calculations.

\end{abstract}

\def\thefootnote{\arabic{footnote}}
\setcounter{footnote}{0}
\newpage


\section{Introduction}

Particle physics is one of the main driving forces in the
development of computing and data
distribution tools for advanced users. Nowadays
computations in particle physics phenomenology take place in a diversified
software ecosystem. In a broad sense we can speak in terms of two
different  categories: commercial or proprietary software, and software
developed by the
scientific collaborations themselves.

Commercial software is distributed under the terms of a particular
end-user license agreement, which defines how and under which
circumstances the 
software should be deployed and used.
In the field of particle physics phenomenology such agreements are
undertaken by the scientific institutions, which afterwards offer this
software as a service to their researchers. This is the case of the
most common software packages employed in the area, such as 
\mma, {\tt Matlab}, etc. 

Scientific collaborations develop also their own software, often
in open source mode under a copy/left license model. 
In this way researchers can download
this software, use it as it is, or implement modifications to
better solve their particular analysis following a GNU General
Public License style%
\footnote{See
{\tt http://www.gnu.org/software/gsl/manual/html\_node/GNU-General-Public-License.html}
for more details}.

From a technical point of view, most of the codes are developed on
Fortran or C/C++.  
They become very modular, because typically they are the result
of the work of a collaborative team on which each member is in charge
of a particular aspect of the calculation. Software packages evolve with
the necessity of analyzing new data, simulating new 
scenarios at present and future colliders. The evolution implies the
inclusion of new modules, or functions, which call and interconnect
other modules in the code, and/or make external calls to proprietary
software like \mma\ to perform basic calculations.

The knowledge of the collaboration and the basics of the physics approach
often resides in the core parts of the code, which remain almost
unaltered for years, while the
development of the software package takes place to include new
features. The core of the software package acts like a
sort of legacy code. The inclusion of new modules to the software package
needs to be done in such a way that these legacy parts remain
untouched as much as possible, because its modification would affect all
modules already present in ways
sometimes very difficult to disentangle, or to predict. All this
reflects in difficulties when it comes to compile those codes together
with more modern ones. Often there are issues with compilers
which cannot be easily solved and require a very deep insight in the
code to be able to install it.

Some of the codes developed in the framework of scientific collaborations 
are not open-source, and therefore
the sources are closed to external researchers.
This reflects situations of competitiveness between groups, and the
fact that the knowledge of the group often resides in the developed
code, and therefore needs to be protected due to Intellectual Property Rights (IPR).

In such situations only the executable binaries are made externally
available by the collaboration, which poses
limitations on the architecture and operating systems, library
versions, etc.\  on which the codes can be executed.

A further level of integration arises when one needs to deal with complex
workflows. This is a most common scenario in particle physics phenomenology
computations: each step of the calculation requires as input the
output of the previous code in the workflow. Therefore, the
installation of several software packages is unavoidable nowadays when,
for instance, the work concerns simulation, prediction or analysis
of LHC data/phenomenology. The installation of several of
those software packages on the same machine is often not trivial since 
one needs to install potentially conflicting software on the same
machine: different libraries for each of these software packages,
sometimes even different compiler versions, etc.

The scenario described results in practical difficulties for
researchers, which translate into time consuming efforts for software
deployment, up to impossibility of deployment due to software or
architecture restrictions. 

There is a general agreement in the community that 
setting up a proper computing environment is becoming a serious overhead
for the everyday work of 
researchers. It is often the case that they need to deploy locally in their
clusters (or even on their own desktops) all the software
packages required for the calculations, each of them with their
particular idiosyncrasies 
regarding compiler versions, dynamic libraries, etc.\ In this case the
intervention of cluster system managers is also not of much help because a
generic cluster cannot accommodate so many options without disturbing
the work of everyone, or generating an unsustainable work overhead to the
system administrator.

\medskip 

The main idea of this work is to exploit the flexibility of operating system
virtualization techniques to overcome the problems described above. 
We will demonstrate how the already available solutions to deploy
cloud computing services \cite{CLOUD} can simplify
the life of researchers doing phenomenology calculations 
and compare the performance to ``more traditional'' installations.

As will be shown along the article, one obvious solution where
virtualization can help with the problems described above is
the deployment of tailored virtual machines fitting exactly the
requirements of the software to be deployed. This is specially the
case when one deals with deploying pre-compiled binaries.
However, the work described here aims for a more complete solution
going from user authentication and authorization, to automation of code
installation and performance analysis.

We want to remark that virtualization techniques are already widely used in most centres
involved in the European Grid Initiative \cite{EGI} as a fault-tolerant mechanism that simultaneouly
allows to simplify the maintainance and operation of the main grid services. However, those
services remain static from the end users perspective, with litle or no possibility to change,
tune or enhance the execution environments. This work is motivated by the necessity of exploring
a more efficient use of computing resources also at the level of the end user. For this purpose,
we are exploring ways in which a Cloud service could be offered as an alternative to experienced
users for which grid infrastructures are no longer able to satisfy their requirements. A mechanism
for authentication and authorization based on VOMS \cite{VOMS} has been develloped and integrated in our
user service provision model to allow interoperability and a smooth, transparent transition between
grid and IAAS cloud infrastructures.

Code performance using virtualized resources versus performances on
non-virtualized hardware is also a subject of debate. 
Therefore it is interesting to make an efficiency analysis 
for real use cases, including self-developed codes and commercial
software in order to shed some light on the influence on the
performance of the host operating system on virtualized environments.

The hardware employed for all the tests described in the article is a
server with 16GB of RAM 
(well above the demand of the applications) with four Intel Xeon
Processors of the family E3-1260L, 8M of Cache and running at 2,40GHz.
In order to have a meaningfull evaluation, we have disabled 
the power efficiency settings in the BIOS to have the processors
running at maximum constant speed. We have also disabled the Turbo
boost features in the BIOS because it increases the frequency of the 
individual cores depending on the occupancy of the cores, therefore
distorting our measures. For the sake of completeness we have
evaluated as well the influence of enabling the Hyperthreading
features of the individual cores to demonstrate
how virtualization and Hyperthreading together influence the performance
of the codes.

The layout of the article is as follows. In Section \ref{sec:cloud} we
describe the architecture and implementation of the proposed solutions;
Sections \ref{sec:single} and \ref{sec:multi} analyze two different real use
cases, together with the respective performance evaluations. 
The first case focuses on the effects of the ``virtual''
environment on single process runs, whereas the second case deals with the potential speed-up
via MPI parallelization on virtualized hardware.
The last section contains our
conclusions. The very technical details
about authentication and user authorization 
as well as detailed numbers about our comparisons can be found in
the Appendix.


\section{Cloud Testbed and Services}
\label{sec:cloud}

\subsection{OpenStack deployment}

The deployment of cloud computing services requires the installation of
a middleware on top of the operating system (Linux in our case), that enables the
secure and efficient provision of virtualized resources in a computing
infrastructure. There are several open-source middleware
packages available to build clouds, with OpenNebula~\cite{ONE} and
OpenStack~\cite{OPENSTACK} being the most used in the scientific data-centers
of the European Grid Infrastructure.

After an evaluation of both OpenNebula and OpenStack, we have chosen the latter
as middleware for our deployment due to its good support for our hardware
and its modular architecture, which allows it to add new services without
disrupting the already existing ones, and to scale easily by replicating
services. OpenStack has a developer community behind that includes over 180
contributing companies and over 6,000 individual members and its being used in
production infrastructures like the public cloud at RackSpace%
\footnote{See {\tt http://www.rackspace.com/cloud/}.}%
. Being written in Python is also an advantage
since we can profit from our expertise in the language to solve problems and
extend the features of the system.

\begin{figure}[htb!]
\begin{center}
\includegraphics[width=0.80\textwidth]{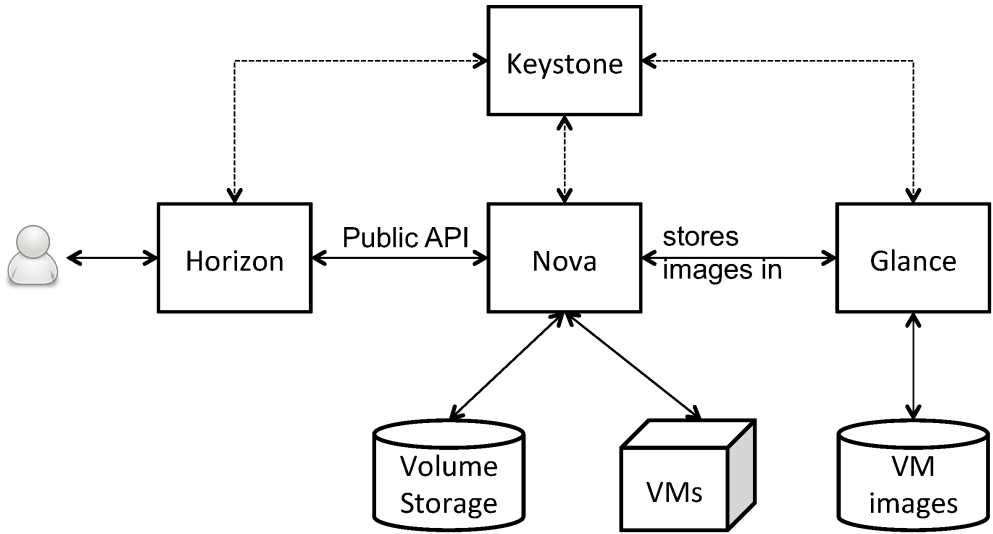}
\caption{OpenStack deployment. Keystone provides authentication for all the
services; Nova provides provisioning of virtual machines and associated
storage; Glance manages the virtual machine images used by Nova; Horizon
provides web-based interface built on top of the public APIs of the services.}
\label{fig:osarch}
\end{center}
\end{figure}

OpenStack is designed as a set of inter-operable services that provide
on-demand resources through public APIs. Our OpenStack
deployment, based on the Essex release (fifth version of OpenStack, released
on April 2012), has the following services, see \reffi{fig:osarch}: 
\begin{itemize}
 \item Keystone (identity service), provides authentication and access control
mechanisms for the rest of components of OpenStack.

 \item Nova (compute service) manages virtual machines and their associated
permanent disks (Volumes in OpenStack terminology). The service provides
an API to start and stop virtual machines at the physical nodes; to assign them
IP addresses for network connectivity; and to create snapshots of
running instances that get saved in the Volume area. The volumes can also be
used as a pluggable disk space to any running virtual machine.

 \item Glance (image management service) provides a catalog and repository for
virtual disk images, which are run by Nova on the physical nodes. 

 \item Horizon, a web-based dashboard that provides a graphical interface to
interact with the services.
\end{itemize}
OpenStack provides also a object storage service but it's not currently used in
our deployment.

Our Nova deployment provides virtual machines using 16 servers
as described in the introduction running Linux  with Xen~\cite{XEN} 4.0.1 as
hypervisor. Volume storage for the virtual machines is provided using two
identical servers with a quad-core Intel Xeon E5606 CPU running at 2.13GHz with
3GB of RAM, 4 TB of raw disk and two 1Gb Ethernet. Glance runs on a server with
similar hardware. 

Users of the infrastructure start virtual machines by selecting one
image from the Glance catalog and the appropriate size (i.e. number of cores,
amount of RAM and disk space) for their computing needs. The available sizes
are designed to fit the physical machine with a maximum of 8 cores and 14GB of
RAM (2GB of RAM are reserved for the physical machine Operating
System, Xen hypervisor and OpenStack services) per machine.

The use of an open-source software allows us to adapt the services to better
suit the needs of a scientific computing environment: we have expanded the
authentication of Keystone to support VOMS and LDAP-based identities as shown
in Appendix~A and we have developed an image contextualization
service with a web interface built on top of Horizon.

\subsection{Image Contextualization}

In an infrastructure as a service cloud, users become the administrators of
their machines. Instead of submitting jobs with their workload to a batch
system where the software is previously configured, they are provided with
virtual machines with no extra help from the resource provider. The
installation and configuration of any additional software must be performed by
the final users. This provides users with flexibility to create tailored
environments for running their software, but requires them to perform tedious
administrative operations that are prone to errors and not of interest
for most users. 

This problem has been partially solved by the CernVM File
System~\cite{CERNVMFS} ---developed to deploy High Energy Physics software on
distributed computing infrastructures--- that provides a read-only file system
for software. However, its centralized design renders it unpractical for
software that changes frequently or is still being developed; it is also limited
to software distribution, which may not be enough for having a working
environment for the researchers. We have developed an image contextualization
service that frees the user from downloading, configuring and installing the
software required for their computations when the virtual machine is
instantiated. This kind of approach does not only provide software installation,
but also allows to customize every other aspect of the machine configuration,
e.g. adding users, mount file-systems (even the CernVM File System) and starting
services.

The service has three main components:  an application catalog
that lists all the available applications; a contextualizer that orchestrates
the whole process and takes care of application dependencies; and a set of
installation scripts that are executed for installation and configuration of
each application. All of them are stored in a git repository at github%
\footnote{See {\tt https://github.com/enolfc/feynapps} for details.}%
. 

The application catalog is a JSON dictionary, where each application is
described with the following fields:
\begin{itemize}
\item 
{\tt app\_name}: human readable application name, for showing it at
user interfaces.

\item
{\tt base\_url}: download URL for the application.

\item 
{\tt file}: name of the file to be downloaded, relative to the {\tt base\_url}.
Applications may be distributed as binaries or source files, the installer
script handles each particular case.

\item 
{\tt dependencies}: list of applications (from the catalog) that need to be
installed before this one.

\item 
{\tt installer}: name of the contextualization script that installs
the application. The contents of this script depend on the characteristics of
the application: it can install additional libraries at the Operating System
level, compile (for applications distributed as source) or simply place
binaries in the correct locations (for applications distributed as binaries).

\item 
{\tt versions}: dictionary containing the different available versions
of the application. Inside this dictionary, there is an entry for each version
where at least a {\tt version\_name} entry specifies a human readable name
for the version. Optionally, it may include any of the fields in the
application description, overriding the default values for the application.
\end{itemize}

The only mandatory fields are the {\tt installer} and {\tt versions}. A sample
entry is shown below. The application name in this case is {\tt FormCalc} and
depends on the {\tt
FeynHiggs} application%
\footnote{
See \refse{sec:codes.and.program.flow} for more details on these codes.}%
. There are two different versions, {\tt 7.0.2} and {\tt
7.4}, with the first one overriding the default value for the {\tt base\_url}:
\begin{verbatim}
"FormCalc": {
  "app_name": "FormCalc",
  "dependencies": [
      "FeynHiggs"
  ],
  "installer": "feyntools.sh",
  "base_url": "http://www.feynarts.de/formcalc/",
  "versions": {
      "7.0.2": {
          "base_url": "https://devel.ifca.es/~enol/feynapps/",
          "app_version": "7.0.2"
      },
      "7.4": {
          "app_version": "7.4"
      }
  }
}
\end{verbatim}

The contextualizer exploits the user-supplied instance meta-data that is
specified at the time of creation of the virtual machine. This is a free form
text that is made available to the running instance through a fixed URL. In our
case, the contextualizer expects to find a JSON dictionary with the
applications to install on the machine. When the virtual machine is started,
the contextualizer fetches the image meta-data and for each application listed
in the JSON dictionary, it downloads the application from the specified URL in
the catalog and executes the installation script. The script contents will
depend on the application to install. It is executed as root user and can
perform any required modifications in the system in order to properly setup the
application (e.g. installation of additional libraries, creation of users,
starting services, etc.). In most cases the script will extract the previously
downloaded application archive and compile it with the virtual machine compiler
and libraries. If the application has any dependencies listed in the catalog,
the contextualizer will install them first, taking care of avoiding duplicated
installations and cyclic dependencies.

The use of a git repository for managing the service provides complete tracking
of the changes in the application catalog and the installation scripts, and
allows researchers to add new applications or enhance the current
installers by submitting pull requests to the repository. It also simplifies
using always up-to-date versions of the tools and catalog at the virtual
machines without having to recreate the virtual machine images by pulling the
latest changes from the repository at instantiation time.

\begin{figure}[htb!]
\begin{center}
\includegraphics[width=0.60\textwidth]{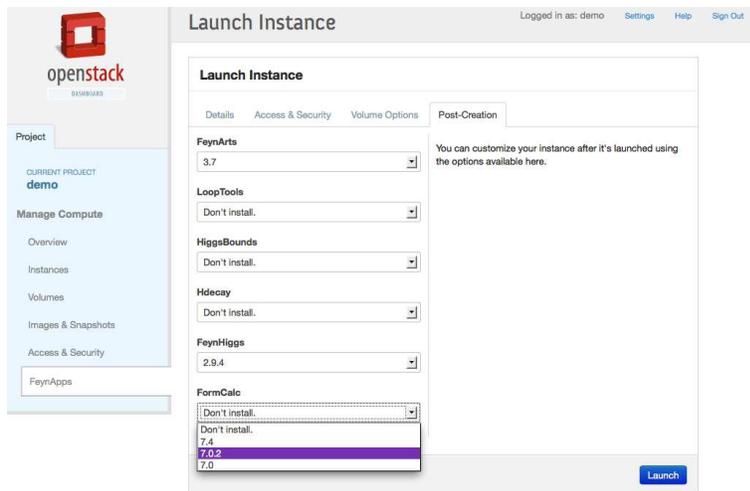}
\caption{Image contextualization panel in Horizon. For each available
application in the catalog, the user can select which version to install.}
\label{fig:horizon}
\end{center}
\end{figure}

To ease the use of the service, we have also extended the OpenStack dashboard to
offer the contextualized instances from a web-based graphical interface.
\reffi{fig:horizon} shows this contextualization panel in horizon. The panel is
a modified version of the instance launch panel, where a new tab includes the
option to select which applications to install. The tab is created on the fly by
reading the application catalog from a local copy of the git repository at the
horizon machine---changes in the application catalog are made available with a
periodic pull of the repository. For each selected application, the panel will
include it in the instance meta-data, which will be used in turn by the
contextualizer to invoke the scripts.

The panel restricts the images that can be instantiated to those that are ready
to start the contextualization on the startup, which are identified in glance
with the property {\tt feynapps} set to {\tt true}. This avoids errors due to
selection of incorrect images and facilitates the addition of new images in the
future without changing the dashboard.



\section{Use Case: single processes on virtual machines}
\label{sec:single}

The first use case analyzed here concerns the evaluation of the decay
properties of (hypothetical) elementary particles. The description of
the underlying physics will be kept at a minimum; more details can be
found in the respective literature.


\subsection{The physics problem}

Nearly all results of high-energy physics results are described with
highest accuracy by the Standard Model (SM) of particle
physics~\cite{sm}. Within this theory it is possible to calculate the
probabilities of elementary particle reactions. A more complicated
theory that tries to go beyond the SM (to answer some questions the SM
cannot properly address) is Supersymmetry (SUSY),
where the most simple realization is the Minimal Supersymmetric Standard
Model (MSSM)~\cite{mssm}. Within this theory all particles of the SM possess
``SUSY partner particles''. The physics problem used in our single-process
example concerns the calculation of the desintegration probabilities 
of one of these SUSY partner particles, 
the so-called ``heaviest neutralino'', which is denoted as $\neu{4}$. 

In the language of the MSSM the two desintegration modes investigated
here are 
\begin{align}
\label{neu4neu1h}
\neu{4} &\to \neu{1} \He~, \\
\neu{4} &\to \chap{1} W^-~.
\label{neu4cha1W}
\end{align}
Here $\neu{1}$ denotes the dark matter particle of the MSSM, $\He$ is a
Higgs boson, $W^-$ is a SM particle responsible for nuclear decay, and
$\chap{1}$ is a corresponding SUSY partner. More details can be found
in \citere{LHCxN}.

The evaluation is split into two parts.
The first part consists of the derivation of analytical formulas that
depend on the free parameters of the model. These parameters are the
masses of the elementary particles as well as various coupling constants
between them. These formulas are derived within 
{\tt Mathematica}~\cite{mathematica} and are subsequently translated into
Fortran code; 
the second part consists of the evaluation of the Fortran
code, see below. Numerical values are given to the free parameters (masses and
couplings) and in this way the desintegration properties for
(\ref{neu4neu1h}), (\ref{neu4cha1W}) are evaluated. In the case of
(\ref{neu4cha1W}) this includes also an additional numerical integration
in four-dimensional space-time, which is performed by the Fortran code.
However, no qualitative differences have been observed, and we will
concentrate solely on process (\ref{neu4neu1h}) in the following.


\newcommand{\mapa}{\mma\ part}
\newcommand{\fopa}{Fortran part}

\subsection{The computer codes and program flow}
\label{sec:codes.and.program.flow}

In the following we give a very brief description of the computer codes
involved in our analysis. Details are not relevant for the
comparison of the different implementations. However, it should be noted
that the codes involved are standard tools in the world of
high-energy physics phenomenology and can be regarded as representative
cases, permitting a valid comparison of their implementation.

\bigskip
The first part of the evaluation is done within 
{\tt Mathematica}~\cite{mathematica} and consequently will be
called ``\mapa'' in the following. It uses several programs
developed 
for the evaluation of the phenomenology of the SM and MSSM. The
corresponding codes are

\begin{itemize}

\item
{\tt FeynArts}~\cite{feynarts}: this {\tt Mathematica} based 
code constructs the ``Feynman diagrams'' and ``amplitudes'' 
that describe the particle decay processes
(\ref{neu4neu1h}) and (\ref{neu4cha1W}). 
This code has been established as a standard tool in high-energy
physics over the last two decades~\cite{feynarts}, as can be seen in the
more than 600 use cases documented~\cite{fa-inspire}.

\item
{\tt FormCalc}~\cite{formcalc}: this {\tt Mathematica} based code
takes the ``amplitudes'' constructed by {\tt FeynArts} and
transforms them into analytical formulas in Fortran.
For intermediate evaluations, {\tt FormCalc} also requires the
installation/use of {\tt Form}~\cite{form}, which is distributed as part
of the {\tt FormCalc} package. 
{\tt FormCalc} is the standard tool to further evaluate {\tt FeynArts}
output, with more than 700 use cases documented~\cite{fc-inspire}.

\item
{\tt LoopTools}~\cite{formcalc}: this Fortran based code provides
four-dimensional (space-time) integrals that are required for the
evaluation of the decay properties.
{\tt FeynArts} and {\tt FormCalc} {\em require} {\tt LoopTools},
i.e.\ it can be seen as an integral part of the above described standard
tool package.

\end{itemize}

Not all parameters contained in the analytical formulas are free,
i.e.\ independent parameters. The structure of the SM and the MSSM fixes
several of the parameters in terms of the others. 
At least one additional code is required to evaluate the dependent
parameters in terms of the others, 

\begin{itemize}

\item
{\tt FeynHiggs}~\cite{feynhiggs}: this Fortran based code provides the
predictions of the Higgs particles (such as $\He$ in \refeq{neu4neu1h})
in the MSSM.
The code has widely been used for experimental and theoretical MSSM
Higgs analyses at LEP, the Tevatron and the LHC. For the latter it has
been adopted as the standard tool for the MSSM Higgs predictions by the
``LHC Higgs Cross Section Working Group''~\cite{lhchxswg,lhchxswg-mssm}.

\end{itemize}

The program flow of the \mapa\ is as follows.
A steering code in \mma\ calls \fa\ and innitates the analytical
evaluation of the decay properties of reaction (\ref{neu4neu1h}) or 
(\ref{neu4cha1W}). In the second step the steering code calls \fc\ for
further evaluation. After the analytical result within \mma\ has been
derived, \fc\ generates a Fortran code that allows for the numerical
evaluation of the results. The code \lt\ is linked to this Fortran code.
Similarly, also \fh\ FeynHiggs is linked to this Fortran code. The creation of the
Fortran code defines the end of the Mathematica part.
The results of these analytical evaluations for the particle processes
under investigations as well as for many similar cases (which used the
same set of codes) have been
verified to give reliable predictions~\cite{moreexamples}.

\begin{figure}[htb!]
\begin{center}
\includegraphics[width=0.80\textwidth]{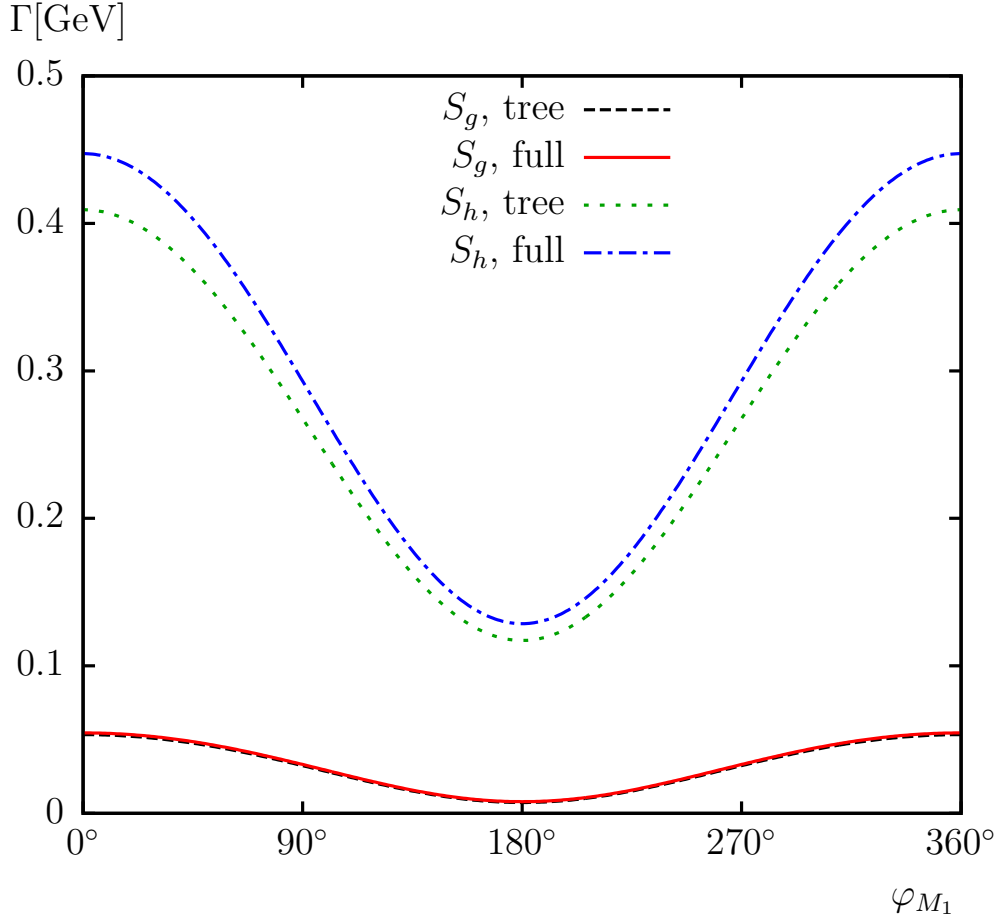}
\caption{
Example output of the evaluation of the properties of decay 
(\ref{neu4neu1h}) for the two parameter configurations, $S_g$ and
$S_h$, in the approximation (``tree'') and the more precise way
(``full'') as a function of $\phiMe$~\cite{LHCxN}.
The decay property $\Ga$ is given in its natural units ``GeV'' (Giga
electron Volt).
}
\label{fig:Gamma}
\end{center}
\vspace{2em}
\end{figure}

\bigskip
The second part of the evaluation is based on Fortran and consequently
will be denoted as ``\fopa'' in the following.
It consists of the execution of the Fortran code created in the
\mapa. One parameter of the model is scanned in a
certain interval, whereas all other parameters of the model are kept
fixed. The calculation of the decay properties are performed for each
value of the varied parameter. To be definite, in our numerical examples
we have varied the complex phase of one of the free
parameters, $\phiMe$, between $0^\circ$ and $360^\circ$ in steps of one
degree. In each of the 361 steps two parameter configurations are
evaluated. Thus, in total the \fopa\ performs 722 evaluations of the
decay properties. As a physics side remark, the results are
evaluated automatically in an approximate way (called ``tree'') and in
a more precise way (called ``full''). 
The results of the \fopa\ are written
into an ASCII file. As an example of this calculation we show in 
\reffi{fig:Gamma} the results for the decay (\ref{neu4neu1h}) for the
two parameter configurations called $S_g$ and $S_h$ (both, ``tree''
and ``full'') as a function of the parameter that is varied, $\phiMe$. 
More details about the physics can be found in \citere{LHCxN}.

\subsection{Performance analysis}

We have measured the performance of the calculation of decay
processes (\ref{neu4neu1h}) and
(\ref{neu4cha1W}) in a virtualized environment.

Our set-up consisted on instantiating virtual machines as described in
\refse{sec:cloud}, including the necessary computational 
packages among them \mma, \fa, \fc, \fh, see above.

Since the nature of the codes is quite different, the computational time
has been measured separately for the \mapa\ of the computation, 
and for the \fopa\ of the  code which involves
basically Floating Point computing (i.e. without the load on file handling
and input/ouput).

In order to fix our notation we introduce the following
abbreviations: 

\begin{itemize}
\item $S_{HT, nHT}(c)$ denotes a virtual machine consisting on $c$
cores and 2GB of RAM.
\item $M_{HT, nHT}(c)$ denotes a virtual machine consisting on $c$
  cores and 4GB of RAM.
\item $L_{HT, nHT}(c)$ denotes a virtual machine consisting on $c$
  cores and 7GB of RAM. 
\item $XL_{HT,nHT}(c)$ denotes a virtual or physical machine with $c$
  cores and 14GB of RAM. 
\end{itemize}

The subscripts $HT$ and $nHT$ refer to Hyperthreading enabled or
disabled on the virtual machine, respectively.
For instance, $M_{HT}(2)$ denotes a virtual machine
with two physical cores, Hyperthreading enabled (i.e. 4 logical cores)
and 4GB of RAM.


\subsubsection{Single process on multicore virtual machines}

\begin{figure}[htb!]
\begin{center}
\includegraphics[width=0.80\textwidth]{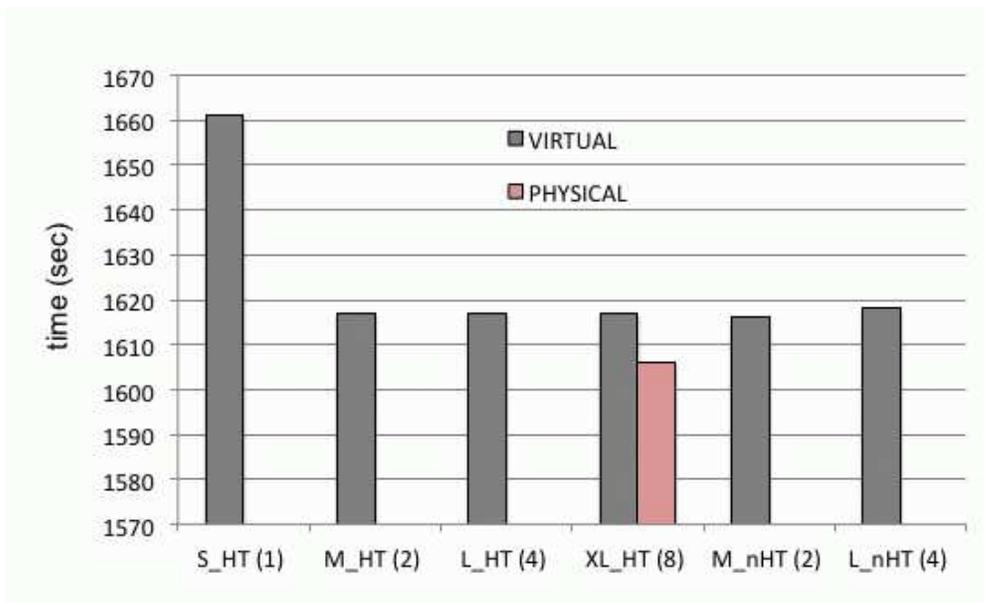}
\caption{Execution time in seconds of the \mapa.
  One single process has been started on the
  different virtual machines configurations. The execution time
on the equivalent physical machine has been included for comparison
for $XL_{HT}(8)$.
The corresponding detailed numbers can be found in \reftas{tab:single} --
\ref{tab:singler}. The scale of the y-axis has been blown to make the
differences visible to the eye.}
\label{fig:one}
\end{center}
\end{figure}

In our first test we submit a single process to the system (regardless
of how many cores are available).
We plot in \reffi{fig:one} the time that only the \mapa\ of the code takes, as a 
function of the configuration of the machine employed. Time measurements
were taken using the GNU time command, that displays information about the
resources used by a process as collected by the Operating System.

As we see the \mapa\ is hardly affected by the
size of the machine, once the virtual machine large enough. The effect
observed with $S_{HT} (1)$ is an overhead due to the extra
work that the only core needs to do to handle both, Mathematica 
and the guest Operating System. Hyperthreading is not enough to
overcome the penalty in performance if only one core is used. 
However when more than one core
is available one can see a constant performance regardless of the
size of the virtual machine, and also regardless or whether
Hyperthreading is enabled or not.

We have also included in this figure the comparison with the time it
takes on the $XL_{HT}(8)$ machine without virtualization, what is
called the ``physical machine''. We see the physical machine is only
slightly faster, 
about a $1\%$. The degradation of performance in this case is therefore
minimal. A more detailed comparison of virtual and physical machines
can be found below.

\begin{figure}[htb!]
\begin{center}
\includegraphics[width=0.80\textwidth]{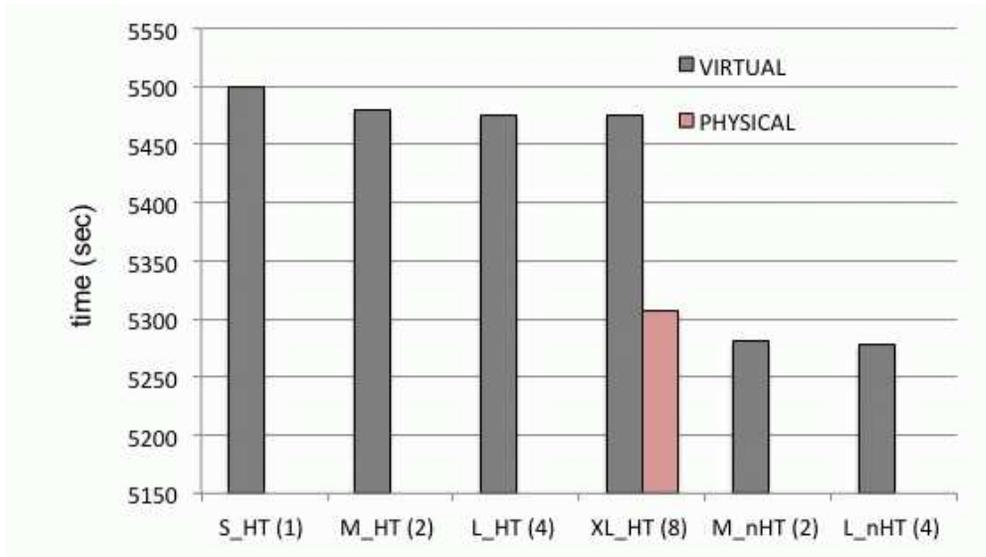}
\caption{Execution time in seconds of the \fopa.
  One single process has been started on the
  different virtual machines configurations. The execution time
on the equivalent physical machine has been included for comparison
for $XL_{HT}(8)$.
The corresponding detailed numbers can be found in \reftas{tab:single} --
\ref{tab:singler}. The scale of the y-axis has been blown to make the
differences visible to the eye. }
\label{fig:two}
\end{center}
\end{figure}

Results turn out qualitatively different in the 
analysis of the \fopa\ of the code, as can be seen in \reffi{fig:two}.
This part is dominated by Floating Point calculations and few
input/output or file handling, 
The first difference we see already at the smaller machines, where we
do not observe anymore overheads due to the size of the virtual machine.
The second difference to the \mapa\ of the code 
is that enabling the Hyperthreading does imply a penalty
on performance on the order of a $4\%$. This is to be expected on
general grounds due to the performance caveats induced by Hyperthreading on
floating-point dominated applications, coming from the fact that the
cores are not physical but logical, and the FPU unit is the same physical one
for the two logical cores.

As for the comparison with the physical machine without virtualization,
again shown for $XL_{HT}(8)$,
we see that virtualization has degraded performance by about a $3\%$
which is still a very small impact. Thus the influence of the host
operating system is very small in low load situations.

For both parts of the evaluation, the \mapa\ and the \fopa,
the percentage of system time employed
during the computations is negligible. For the Mathematica dominated
part of the computation it starts at $3\%$ in $S_{HT}(1)$, to decrease
down to a $1,5\%$ in the rest of series.
In the Fortran part it stays constant at about $0.2\%$.


\subsubsection{Multiple simultaneous processes on multicore 
virtual machines}

In this section we investigate the behavior of the performance in
virtual machines under high load circumstances. For that we use a
machine with 4 physical 
cores, Hyperthreading enabled, thus 8 logical cores. 

To fix the notation we have adapted the previous
definition as follows (in this test Hyperthreading
is always enabled, therefore we drop the subscript for simplicity)

\begin{itemize}
\item $M(c/p)$ denotes a virtual machine consisting on $c$
  cores and 4GB of RAM and $p$ concurrent processes running.
\item $L(c/p)$ denotes a virtual machine consisting on $c$
  cores and 7GB of RAM and $p$ concurrent processes running.
\item $XL(c/p)$ denotes a virtual or physical machine with $c$ cores and
  14GB of RAM and $p$ concurrent processes running.
\end{itemize}

The test was performed as follows. First we instantiate a virtual
machine with a number of logical cores $c$. 
Then we start from $p=1$ up to $p=c$ simultaneous processes in order
to fill all the logical cores available, and measure how long each
of the simultaneous processes takes to complete. 
Since not all the simultaneous processes take the same time to
complete, we have taken the time of the slowest one for the plots.
Conservatively speaking,  this is the real time that the user
would have to wait. The difference between the maximum and minimum
times is not significative for our analysis (see \reftas{tab:multi.vht},
\ref{tab:multi.r} in Appendix~B for more
details on actual times).

\begin{figure}[htb!]
\begin{center}
\includegraphics[width=0.80\textwidth]{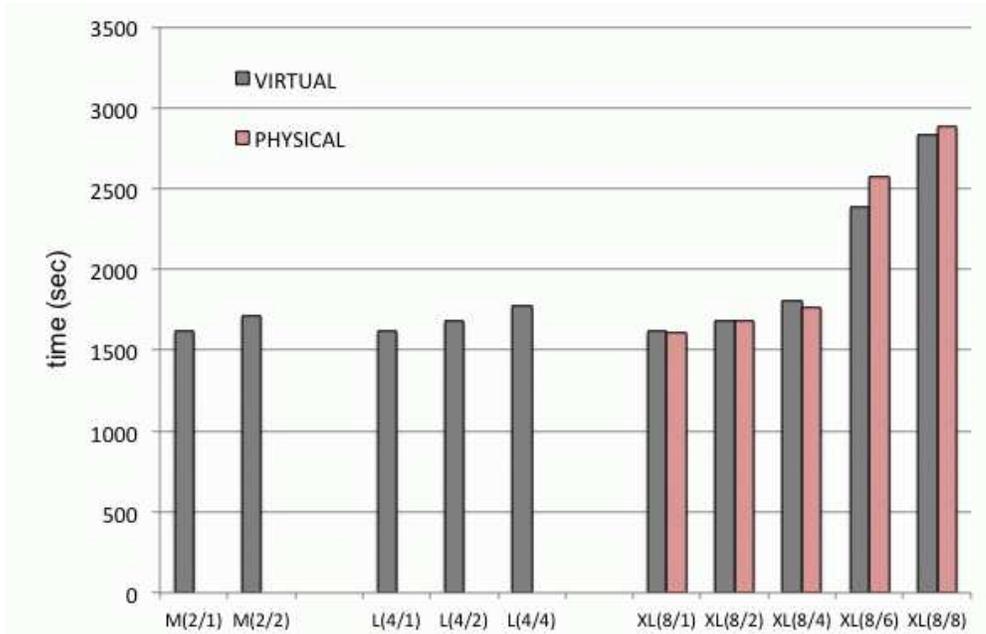}
\caption{Execution time in seconds of the parts of the calculation
  involving Mathematica. The execution time
on the equivalent physical machine has been included for
comparison.
The corresponding detailed numbers can be found in \reftas{tab:multi.vht},
\ref{tab:multi.r}.}
\label{fig:three}
\end{center}
\end{figure}

In \reffi{fig:three} we plot the execution time in seconds of the
\mapa\ of the code for the $M$, $L$ and $XL$ machines with various
number of processes as described above. In the $XL$ case, for
comparison, we also show the execution time in the physical machine.
The first observation is that
the degradation on the performance appears only when we load the
system with more processes than the existing physical cores
(i.e.\ more than $4$). Thus we conclude that this is not an effect of
virtualization, but rather of Hyperthreading. 
In the comparison of the virtual and the physical machines, shown for
$XL(8/n)$ in \reffi{fig:three}, one can see that the 
virtualization does not really imply a penalty on the performance.

An interesting effect in this comparison can be observed 
when submitting $p=6$ or more simultaneous
processes. Against intuition the physical machine execution
time is larger than the virtual machine execution time. 
This fact can only be explained if the virtualized operating system
manages to handle better the threads than the normal operating
system, which relies only Hyperthreading to distribute the system load.

\begin{figure}[htb!]
\begin{center}
\includegraphics[width=0.80\textwidth]{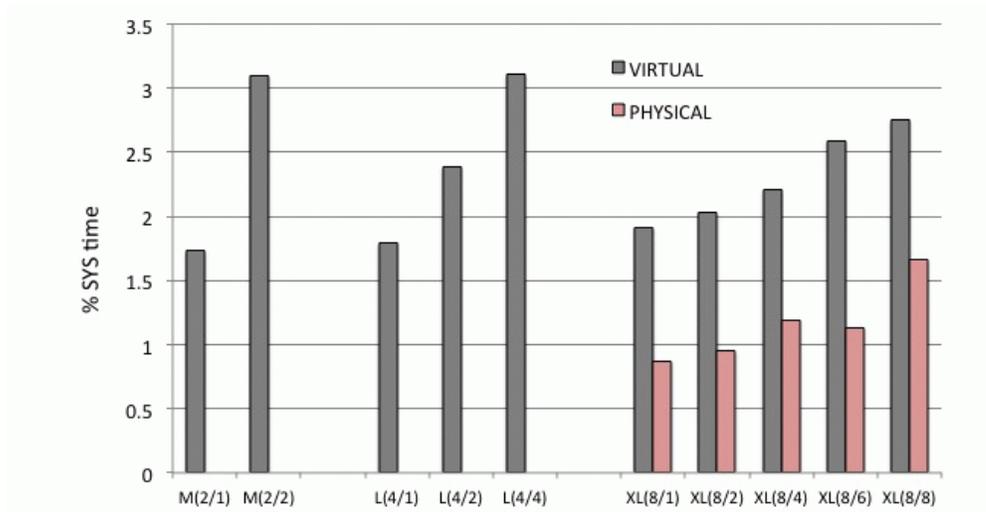}
\caption{Percentage of system time employed by the virtual machine in
  the \mapa. The same percentage
on the equivalent physical machine has been included for comparison
in the $XL$ case.
The corresponding detailed numbers can be found in 
\reftas{tab:multi.vht},~\ref{tab:multi.r}.}
\label{fig:four}
\end{center}
\vspace{1em}
\end{figure}

To investigate this effect we plot in \reffi{fig:four} the
percentage of system time which the operating systems employed on the
runs. We can see how at $XL(8/6)$ the physical machine does spend less sytem
time than expected, and indeed, it is not managing the load of the 6
processes on the 8 logical cores in the most efficient way. In this
case the spread in execution time between the fastest and the slowest 
processor is very large (2572 seconds versus 1899 seconds, where
the latter is faster than the fastest time on the virtual machine,
2359 seconds),  
see \reftas{tab:multi.vht}, \ref{tab:multi.r} in Appendix~B.

\begin{figure}[htb!]
\begin{center}
\includegraphics[width=0.80\textwidth]{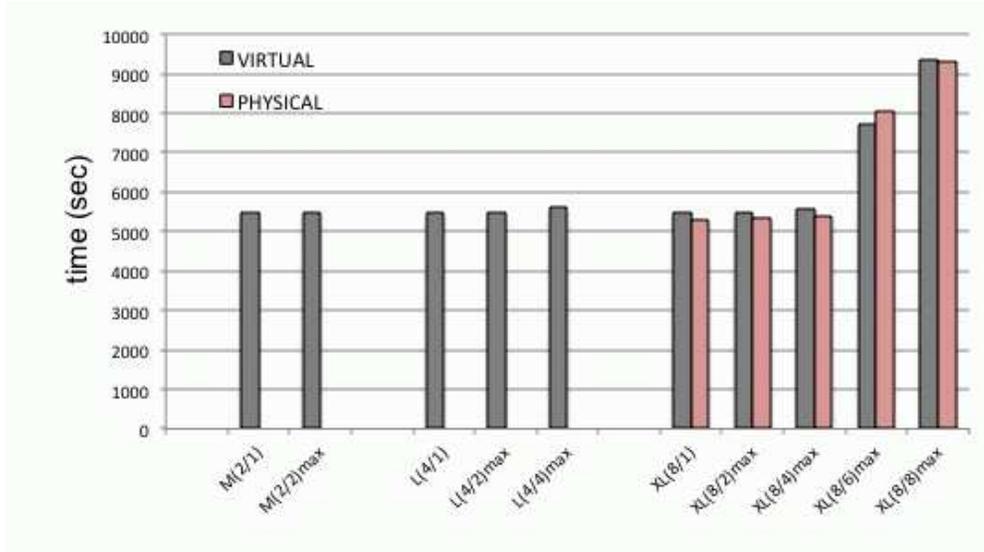}
\caption{Execution time in seconds of the \fopa. The execution time
on the equivalent physical machine has been included for comparison
for the $XL$ case.
The corresponding detailed numbers can be found in \reftas{tab:multi.vht},
\ref{tab:multi.r}.}
\label{fig:five}
\end{center}
\vspace{1em}
\end{figure}

To conclude we plot in \reffi{fig:five} the equivalent
execution times in the Fortran dominated part of the calculation. We
see that essentially the same pattern of behavior reproduces:
the load of the machines have a sizable effect on the execution
 time only for more than 4~simultaneous processes, and the virtual and
 physical machines show negligible differences.

We also measured the memory consumption of the applications to ensure that
swapping had no effects on the applications' execution. The \mapa\ memory
footprint was collected using {\tt Mathematica} memory management variable {\tt
MaxMemoryUsed[]}; while the \fopa\ footprint was measured with the
Valgrind~\cite{VALGRIND} heap profiler tool. The maximum
memory consumption for the \mapa\ was 691.1~MB. The \fopa\ had a lower
memory consumption with a maximum of 189.1~MB for the compilation of the
resultant codes from {\tt FormCalc} and 36.9~MB for the execution. These values
are well below the minimum 1.75 GB of RAM per core (as in the $XL(8/n)$ case)
available in the virtual machines. The possibility of selecting the size of the
virtual machine upon startup allows users to adapt their virtual
infrastructure to the particular memory requirements of their applications.


\newpage

\section{Use Case: MPI Parallelization}
\label{sec:multi}

The second use case analyzed here concerns a parameter scan as a typical
application in the field of high-energy physics phenomenology. It also
constitutes a perfect example that can be  easily parallelized, see below for
more details.
For each point in the parameter scan an evaluations of Higgs boson properties
that depend on this parameter choice is performed. As in the previous section,
the description of the underlying physics will be kept at a minimum, and more
details can be found in the respective literature. 


\subsection{The physics problem}

Also this physics problem is taken from the MSSM.
This model possesses several free parameters. Since they are unknown,
a typical example of an analysis within this model requires extensive
parameter scans, where the predictions for the LHC phenomenology change
with the set of the scanned parameters. 

After the discovery of a Higgs-like particle at the
LHC~\cite{ATLASdiscovery,CMSdiscovery} the Higgs bosons of the MSSM are
naturally of particular interest. 
The most relevant free parameters of the MSSM in this respect are
\begin{align}
\MA \mbox{~and~} \tb~.
\end{align}
$\MA$ denotes the mass of a Higgs particle in the MSSM, $\be$ is a
``mixing angle'', see \citere{benchmark2} for further details.

A typical question for a choice of parameters is, whether this
particular combination of parameters is experimentally allowed or
forbidden. A parameter combination, in our case a combination of $\MA$
and $\tb$, can result in predictions for the Higgs particles that are in
disagreement with experimental measurements. Such a parameter
combination is 
called ``experimentally excluded''. In the example we are using, two
experimental results are considered. The first are the results from the
LHC experiment itself. The other set are the results from a previous
experiment, called ``LEP''~\cite{LEPHiggs}.


\subsection{The computer codes and program flow}
\label{sec:mpi-codes}

In the following we give a very brief description of the computer codes
involved in this analysis. Details are not relevant for the
comparison of the various levels of parallelization. As in the previous
example, it should be noted that the codes involved constitute standard
tools in the world of 
high-energy physics phenomenology and can be regarded as representative
cases, permitting a valid comparison of their implementation.

The main code that performs the check of a Higgs prediction with results
from the LHC and LEP is 

\begin{itemize}

\item
{\tt HiggsBounds}~\cite{higgsbounds}: 
this Fortran based code takes input for the model
predictions from the user and compares it to the experimental results
that are stored in the form of tables (which form part of the code).
{\tt HiggsBounds} has been established as the standard tool for the
generic applicatino of Higgs exlcusion limits over the last years. It
has been linke to many other high-energy physics standard codes to
facilitate their evaluation~\cite{hb-inspire}.

\end{itemize}

The predictions for the Higgs phenomenology are obtained with the same
code used in the previous section, 

\begin{itemize}

\item
{\tt FeynHiggs}~\cite{feynhiggs}: this Fortran based code provides the
predictions of the Higgs particles in the MSSM
(for more details see the previous section).

\end{itemize}

In our implementation a short steering code (also in Fortran) 
contains the initialization of the parameter scan: two loops over the
scan parameters, $\MA$ and $\tb$, are performed in the ranges 
(omiting physical units), 
\begin{align}
\MA &= 90 \ldots 500~, \non \\
\tb &= 1.1 \ldots 60~, 
\end{align}
with 120 steps in each parameter, resulting in 14400 scan points.
As a physics side remark: the other free parameters are set
to fixed values, in our case according to the $\mhmax$ scenario
described in \citere{benchmark2}. However, details are not relevant for
our analysis.

\begin{figure}[htb!]
\begin{center}
\includegraphics[width=0.80\textwidth]{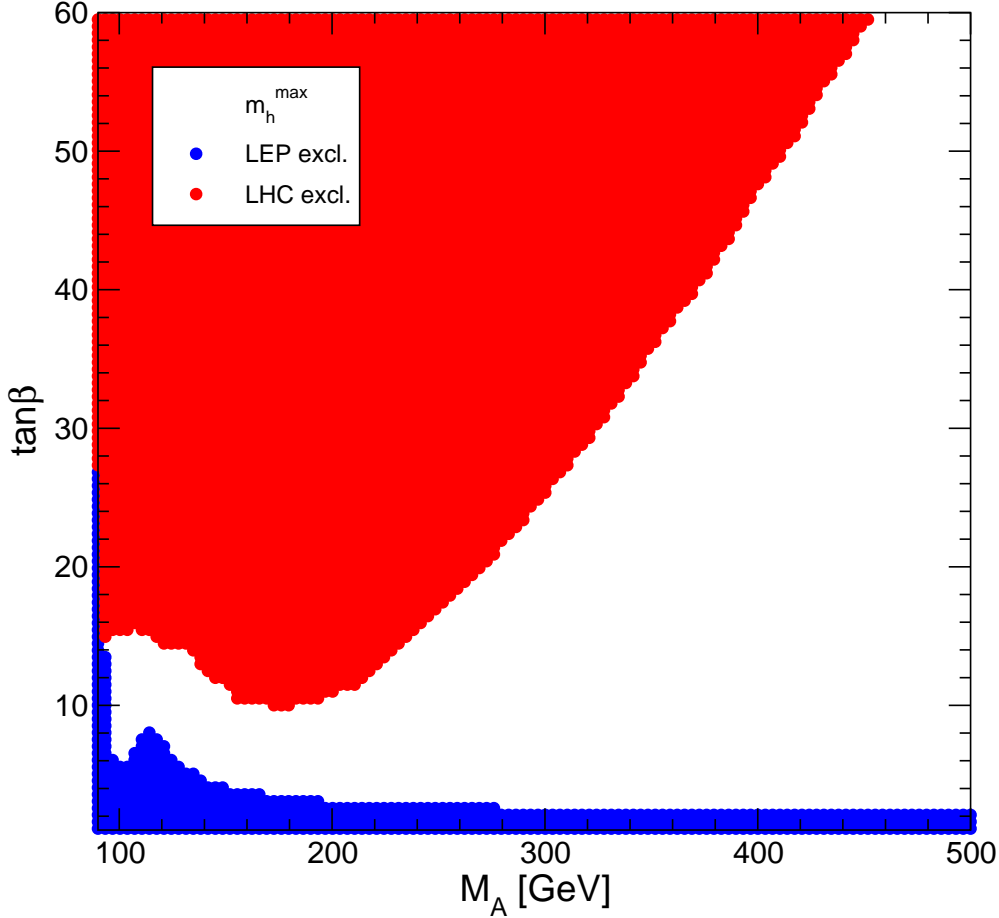}
\caption{
Example output of the MSSM scan in the two free parameters $\MA$ and
$\tb$. The parameter $\MA$ is given in its natural units ``GeV'' (Giga
electron Volt).
}
\label{fig:mhmax}
\end{center}
\end{figure}

The steering code calls the code {\tt HiggsBounds}, handing over the
scan parameters. Internally {\tt HiggsBounds} is linked to 
{\tt FeynHiggs}, again handing over the scan parameters.
{\tt FeynHiggs} performs the prediction of the Higgs phenomenology, and
the results are given back to {\tt HiggsBounds}. With these parameters
the code can now evaluate whether this parameter combination is allowed
or disallowed by existing experimental results. The corresponding
results are stored in a simple ASCII file, where one file contains the
points excluded by the LHC, another file the points excluded by LEP. 
As an example, we show in \reffi{fig:mhmax} the results for this scan in
the two-dimensional $\MA$-$\tb$ plane. Points marked in red, according
to the evaluation with {\tt HiggsBounds}/{\tt FeynHiggs} are in
disagreement with experimental results from the LHC, and blue points are
in disagreement with experimental results from LEP. White points are in
agreement with the currently available experimental results.

\subsection{MPI parallelization}

The parameter scan performed by the code is a typical example of 
an embarrassingly parallel computation, where each parameter evaluation
can be computed 
independently of the others, without requiring any communication between them.
This kind of problems can be easily parallelized by dividing the parameter
space into sets and assign them to each available processor. An
OpenMP~\cite{OPENMP} parallelization was discarded due to the use of non
thread-safe libraries in the code, so we opted for using MPI~\cite{MPI} for
developing the parallel version of the code.

In the parallel version, the steering code in Fortran was modified to 
have a single process that initializes the computation by setting the number
of steps (by default 120 steps in each parameter) and values for the fixed free
parameters and broadcasting all these values to the other processes in the
computation. The parameter space is then divided equally
among all processes, which perform the evaluation and write
their partial results to independent files without any further communication
between processes. Once the computation finishes, the partial results files are
merged into a single file with all results. A master/worker parallelization with
dynamic assignment of the parameters to each worker was not considered because
the execution time per evaluation is almost constant hence there is no need to
balance the work load between the workers.


\subsection{Performance analysis}

We have measured the scalability and performance of the two-dimensional
$\MA$-$\tb$ plane scan described in Section~\ref{sec:mpi-codes} with $14400$
scan points in a virtualized environment. As in the previous case, we have
instantiated the virtual machines using our contextualization mechanism to
install {\tt FeynHiggs} and {\tt
HiggsBounds} packages. The MPI code was compiled with {\tt Open
MPI} v1.2.8~\cite{OPENMPI} as provided in the Operating System distribution.

These tests were performed on virtual machines that use the complete hardware,
with and without Hyperthreading enabled (4 or 8 logical cores
respectively) and the equivalent physical machine with the same number of cores
and RAM to compare the performance without virtualization.

\begin{figure}[htb!]
\begin{center}
\includegraphics[width=0.8\textwidth]{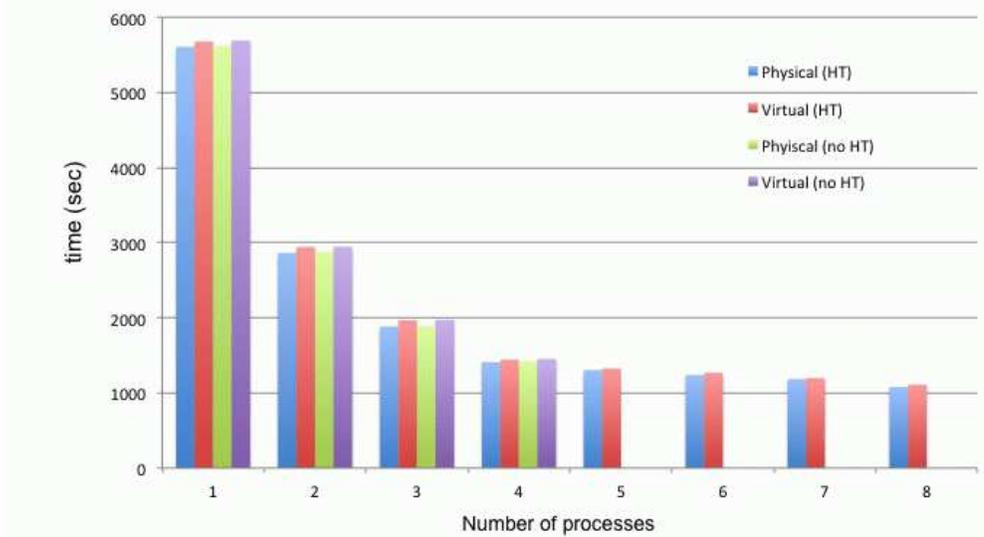}
\caption{Execution time in seconds of the application for different number of
processes, both in Virtual and Physical machines with and without
HyperThreading.}
\label{fig:mpicompare}
\end{center}
\end{figure}

We plot in \reffi{fig:mpicompare} the execution time for the parameter scan
using from 1 (serial version) up to the number of cores available in each
machine. The parallel versions time include also the final merge of the partial
result files.

As we see the performance degradation due to virtualization is
minimal, below $5\%$ for all executions, and the difference in execution time
with and without HyperThreading for the same number of processes is negligible.
The difference between the virtual and physical machine decreases as the number
of processes grows above 4. This effect, also seen in the case of multiple
processes in Section \ref{sec:single}, is due to the use different management
the HyperThreading cores at the virtualized Operating System.

Since there is no communication overhead in the implementation,
the application scales linearly with the number of processes given equally
powerful CPUs. As seen in the plot, the scalability of the application is almost
linear up to $4$ processes (the same number of processes as available physical
cores) and it flattens as the Operating System uses the logical cores provided
by the HyperThreading.



\section{Conclusions}

We have described a new computing environment for particle physics
phenomenology that can easily be translated to other branches of
science. It is based on ``virtual machines'', using the OpenStack
infrastructure. In view of the performance, 
it is necessary to distinguish between two questions: the benefits that
virtualization brings to researchers in terms on accesibility to
computing resources, and the question of code performance and in general
penalties due to the host operating system.

About the first question the setup of OpenStack and the development of
the self-instant-\\
iation mechanism has been clearly appreciated by the
researchers doing this type of computations. 
The solution removes many
of the barriers described in the introduction of this article
regarding complex code installation, machine availability, and
automatization of workflows. 

An additional benefit of this set-up is that OpenStack allows the
user taking snapshots of the virtual machine, which are stored on a
repository, and which the owner of the snapshot can instantiate again
at any moment, recovering the session as they saved it. This is a very
practical feature because it allows 
researchers to ``save'' the current status on the virtual machine, and
continue working at any other moment without blocking the hardware in
the mean time.

The second question is performance. We have analyzed a set of
representative codes in the area of particle physics phenomenology,
so that our results can be extrapolated to similar 
codes in the area. The results are very positive, 
as no excesive penalty due to virtualization can be
observed. At most we observe degradations in performance on the order of
$3\%$ for the parts of the codes dominated by Floating Point
Calculations. For other calculations the degradation was even less.
We have furthermore analyzed the 
influence of system time in the virtual machines. We found that the
virtualization has no significant impact on the system time.

Evidently, the possibility of accessing resources in a
more flexible way, the time that researchers spare using the new
environment on software configuration compensates largely the usage of
virtualized resources for the codes under investigation.


\subsection*{Acknowledgements}

I.C.\ and E.F.\ thank the European Commission funding via EGI-InSPIRE 
Grant Contract number RI-261323, EMI Grant Contract number RI-261611
and FPA-2008--01732.
The work of S.H.\ was partially supported by CICYT (grant FPA
2010--22163-C02-01).
S.H.\ and F.v.d.P.\ were supported in part by 
the Spanish MICINN's Consolider-Ingenio 2010 Programme under grant
MultiDark CSD2009-00064 and the project ``Computacion Avanzada en
materiales y fenomenos de transporte'' Grant number
MEC-FIS-2009-12648-C03-02. 
G.B.\ would like to thank to the Portuguese Foundation for Science and
Technology under the context of the Ciencia 2008 program jointly funded by the
European Social Fund and by MCTES national funds - through POPH - NSRF-Type
4.2.


\section*{Appendix A: User access mechanisms}

In this Appendix we describe the user access mechanisms that we have
implemented. 

\subsection*{A.I Authentication}
\label{sec:auth}

Keystone performs the validation of user's credentials (username/password)
using a configurable back-end, which also stores all the data and associated
meta-data about users, \emph{tenants} (groups in OpenStack terminology) and
roles associated to each user. There are four back-ends provided in the default
OpenStack distribution: Key-Value Store (KVS), which provides an interface
to other stores that can support primary key lookups (like a hash table);
SQL, that stores data in a persistent database using SQLAlchemy; LDAP, that
uses a LDAP server where users and tenants are stored in separate subtrees; and
PAM, for simple mapping of local system users to OpenStack. The typical
deployment uses the SQL backend.

Additional authentication mechanisms, e.g. those not based on username/password
credentials, can be implemented using the Pluggable Authentication Handler
(PAH) mechanism. A pluggable authentication handler analyzes each user request
and, if it's able to handle it, authenticates the user before the default
back-end is executed.

Scientific computing and data centers provide services to scientific collaborations that often
go beyond a single institution, creating federation of resources where users
can access to the different providers using a single identity. 

Grid infrastructures base the authentication and authorization of users on X.509
certificates and Virtual Organizations (VO). A Virtual Organization comprises a
dynamic set of individuals (or institutions) defined around a set of
resource-sharing rules and conditions. The current pan-european grid
infrastructure use the Virtual Organization Management System (VOMS) for
managing the users and resources within each VO. VOMS provides signed assertions
regarding the attributes of a user belonging to a VO, thus enabling providers to
trust these assertions and to define access rules to their resources based on
the attributes. While a broad community of users are already familiarized with
these authentication mechanisms, the use of X.509 certificates and proxies is
considered to be one of the main barriers for new users and communities.

There are other mechanisms to provide identity federation across multiple
providers that do not require the use of certificates, but these are poorly
supported by the current scientific collaborations for accessing computing
resources. LDAP can also be used to support federated
authentication. LDAP is a well-known solution for authentication and it is
widely used within the scientific data centers. However, the LDAP back-end
included Openstack enforces a particular schema that does not fit most of
existing deployments.

We have extended the authentication capabilities of Keystone with two new
Pluggable Authentication Handlers, one supporting VOMS and another supporting
LDAP authentication with arbitrary schema for storing user information. These
modules enable the creation of a federated cloud infrastructure where users
have a single identity across the different resource providers. The VOMS module
enables the re-use of grid identity systems and leverages from existing
experience of the resource providers, while the LDAP module enables the creation
of simple federation without the inconvenience of the X.509 certificates for new
users and communities.

\subsection*{A.II VOMS Authentication}

\begin{figure}[htb!]
\begin{center}
\includegraphics[width=0.75\textwidth]{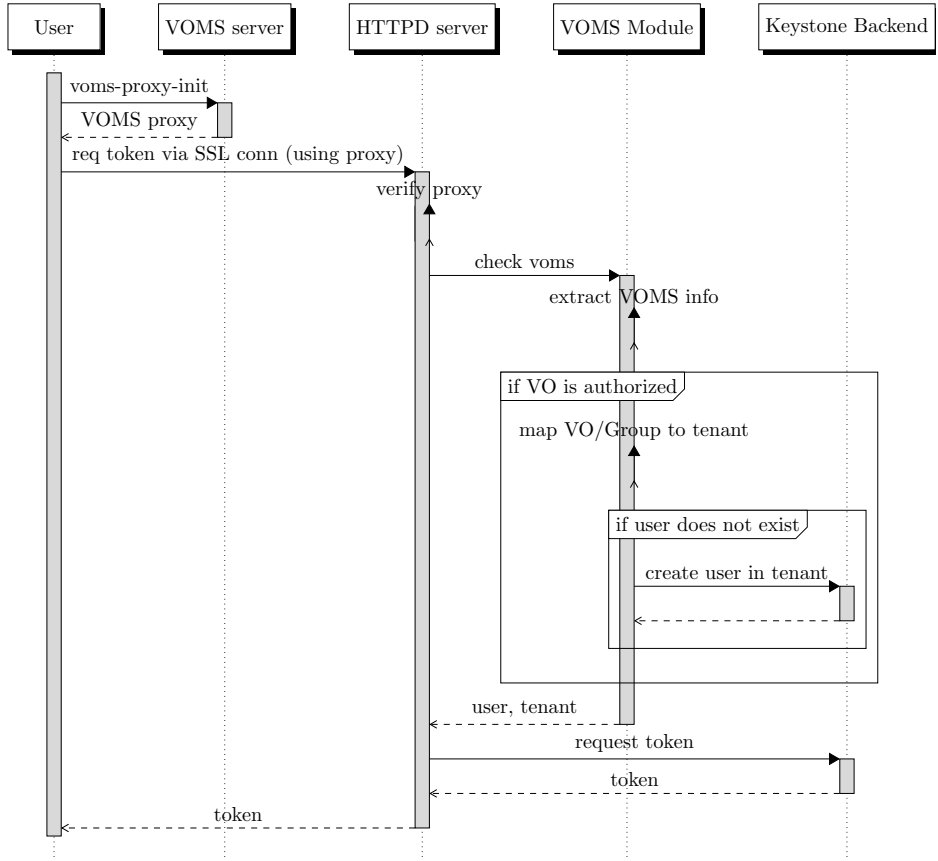}
\caption{VOMS Authentication sequence diagaram.}
\label{fig:ksvoms}
\end{center}
\end{figure}

The VOMS authentication module is implemented as a Pluggable Authentication
Handler in Keystone and executed as a WSGI~\cite{WSGI} of an httpd server
enabled to use OpenSSL and configured to accept proxy certificates (VOMS
assertions are included in proxy certificates). \reffi{fig:ksvoms} shows the
call sequence for the authentication and authorization of a user using VOMS.

Prior to the authentication, the user creates a proxy by contacting the VOMS
server. This proxy includes the distinguished name (DN) of the user and a set
of attributes related to the VO (group and roles associated with the user). The
authentication in Keystone is performed by requesting a token to the
server, if successful, Keystone will return a token which is used for
any subsequent calls to the other OpenStack services. In our implementation,
the user authenticates against the httpd server with the VOMS proxy, and the
server, after validation, includes the SSL information in the request
environment. This information reaches the VOMS module that will authorize the
request checking if the proxy is valid and if the VO is allowed in the server. 

Once the proxy is considered valid and allowed, the module maps VO attributes
included in the proxy to a local OpenStack tenant using a configuration file.
The user DN in the proxy is used as user name in Keystone. The mapped
local tenant must exist in advance for a user to be authenticated. The VOMS
module can automatically create the user in Keystone if enabled in
configuration. This allows to easily establish access policies based only in the
membership to a given VO, instead of giving access to individuals.
Once a user has been granted access, the administrator can manage it as with any
other user in the Keystone back-end (i.e. disable/enable, grant/revoke roles,
etc.).

\subsection*{A.III LDAP Authentication}

The LDAP authentication modules takes profit from the authentication features
of the apache httpd server. In this case the authentication phase is
delegated to the server which can authenticate users against LDAP
with arbitrary schemas. This way, we use a reliable and tested code that is
used in production systems and is actively maintained by the apache
developers. Moreover, we avoid the introduction of security risks by
minimizing the code the deals with the authentication. Once the server has
authenticated the request, it will pass the user name through the
environment of the WSGI module. 

\begin{figure}[htb!]
\begin{center}
\includegraphics[width=0.75\textwidth]{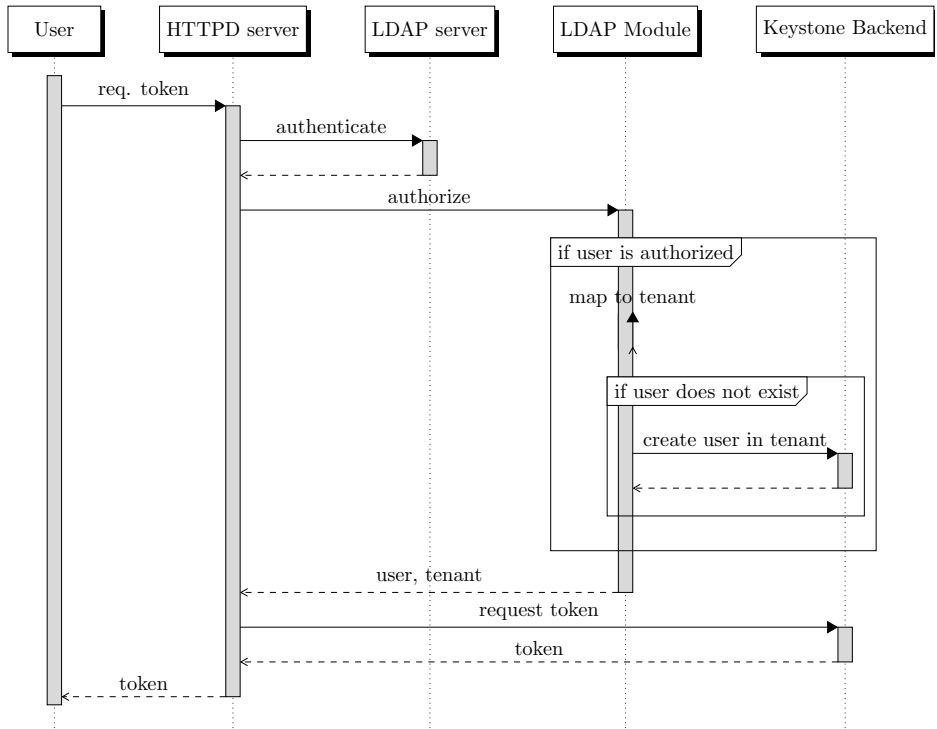}
\caption{LDAP Authentication sequence diagaram.}
\label{fig:ksldap}
\end{center}
\end{figure}

The LDAP module uses a configuration file that specifies using regular
expressions which users are allowed and to what tenants they should be mapped
in the system. If the user is authorized, the user can be
created automatically in Keystone as in the VOMS case. Finally, the token is
returned to the user. Figure \ref{fig:ksldap} shows the sequence diagram for
this process.

\newpage

\section*{Appendix B: Detailed computation times}


In this Appendix we give detailed numbers on the execution times of the
analysis in \refse{sec:single}. The notations are the same as in that
section. 

\begin{table}[ht!]
\renewcommand{\arraystretch}{1.5}
\BC
{\small 
\begin{tabular}{|l|r|r|r|r|r|r|r|r|r|r|r|}
\hline
machine &  \multicolumn{3}{c|}{Math} 
      &  \multicolumn{3}{c|}{Fortran} 
      &  \multicolumn{3}{c|}{Total} 
\\ 
\cline{2-10}
(cores)
& real  & user & sys
& real  & user & sys
& real  & user & sys
\\ \hline
$S_{HT}(1)$
& 1661.06 & 1601.87 & 48.38
& 5500.06 & 5466.54 & 11.07
& 7161.24 & 7068.43 & 59.47
\\ \hline
$M_{HT}(2)$ 
& 1617.06 & 1613.56 & 28.37
& 5480.03 & 5465.36 & 12.20
& 7097.20 & 7078.99 & 40.58
\\ \hline
$L_{HT}(4)$ 
& 1617.05 & 1615.18 & 29.38
& 5475.64 & 5461.08 & 12.28
& 7092.80 & 7076.33 & 41.68
\\ \hline
$XL_{HT}(8)$ 
& 1617.19 & 1615.86 & 31.00
& 5475.95 & 5461.20 & 12.48
& 7093.25 & 7077.12 & 43.50
\\ \hline
\end{tabular}
}
\caption{
Computation time (sec) of virtual machines with hyperthreading (HT), 
divided in \mapa, \fopa, and total time.
Computation time is divided in real, user, sys, 
denoting respectively real time, processor time for computation, 
and system time.
}
\label{tab:single}
\EC
\renewcommand{\arraystretch}{1.0}
\vspace{-2em}
\end{table}

\begin{table}[ht!]
\renewcommand{\arraystretch}{1.2}
\BC
{
\begin{tabular}{|l|r|r|r|r|r|r|r|r|r|r|r|}
\hline
machine &  \multicolumn{3}{c|}{Math} 
      &  \multicolumn{3}{c|}{Fortran} 
\\ 
\cline{2-7}
(cores)
& real  & user & sys
& real  & user & sys
\\ \hline
$M_{nHT}(2)$
& 1616.78 & 1613.12 & 28.20
& 5281.21 & 5266.46 & 11.69
\\ \hline
$L_{nHT}(4)$
& 1618.9  & 1613.9  & 33.8 
& 5278.9  & 5264.0  & 12.6 
\\ \hline
\end{tabular}
}
\caption{
Computation time (sec) of virtual machines without hyperthreading (nHT). 
}
\label{tab:singlenht}
\EC
\vspace{-0.5em}
\renewcommand{\arraystretch}{1.0}
\end{table}

\begin{table}[ht!]
\renewcommand{\arraystretch}{1.2}
\BC
{
\begin{tabular}{|l|r|r|r|r|r|r|r|r|r|r|r|}
\hline
machine &  \multicolumn{3}{c|}{Math} 
      &  \multicolumn{3}{c|}{Fortran} 
\\ 
\cline{2-7}
(cores)
& real  & user & sys
& real  & user & sys
\\ \hline
$R_{HT}(8)$ 
& 1605.1 
& 1584.8 
& 13.9 
& 5306.9 
& 5290.8 
& 6.6 
\\ \hline
\end{tabular}
}
\caption{
Computation time (sec) of physical machine $R$ with
hyperthreading (HT). 
}
\label{tab:singler}
\EC
\vspace{-0.5em}
\renewcommand{\arraystretch}{1.0}
\end{table}

\newpage

\begin{table}[ht!]
\renewcommand{\arraystretch}{1.4}
\BC
{
\begin{tabular}{|l|r|r|r|r|r|r|r|r|r|r|r|}
\hline
machine &  \multicolumn{3}{c|}{Math} 
      &  \multicolumn{3}{c|}{Fortran} 
\\ 
\cline{2-7}
(cores/proc.)
& real & user  & sys
& real & user  & sys
\\ \hline
$M_{HT}(2/1)$ 
& 1617.1  & 1613.6  & 28.4 
& 5480.0  & 5465.4  & 12.2 
\\ \hline
$M_{HT}(2/2)$ max 
& 1708.9  & 1647.2  & 52.6 
& 5500.7  & 5472.7  & 12.2 
\\ \hline
$M_{HT}(2/2)$ min
& 1713.4  & 1650.6  & 53.4 
& 5493.6  & 5469.7  & 12.2 
\\ \hline\hline
$L_{HT}(4/1)$ 
& 1617.1  & 1615.2  & 29.4 
& 5475.6  & 5461.1  & 12.3 
\\ \hline
$L_{HT}(4/2)$ max
& 1678.6  & 1653.2  & 39.5 
& 5488.0  & 5473.3  & 12.4 
\\ \hline
$L_{HT}(4/2)$ min
& 1672.4  & 1656.4  & 36.1 
& 5491.0  & 5476.0  & 12.6 
\\ \hline
$L_{HT}(4/4)$ max
& 1771.1  & 1709.0  & 54.5 
& 5602.4  & 5580.7  & 12.3 
\\ \hline
$L_{HT}(4/4)$ min
& 1775.3  & 1711.8  & 53.8 
& 5511.9  & 5489.8  & 12.7 
\\ \hline\hline
$XL_{HT}(8/1)$ 
& 1617.2  & 1615.9  & 31.0 
& 5476.0  & 5461.2  & 12.5 
\\ \hline
$XL_{HT}(8/2)$ max
& 1678.4  & 1671.2  & 34.4 
& 5492.4  & 5477.1  & 13.1 
\\ \hline
$XL_{HT}(8/2)$ min
& 1676.8  & 1668.6  & 35.3 
& 5493.4  & 5478.2  & 13.0 
\\ \hline
$XL_{HT}(8/4)$ max
& 1807.0  & 1790.9  & 39.7 
& 5566.1  & 5549.4  & 13.9 
\\ \hline
$XL_{HT}(8/4)$ min
& 1809.6  & 1786.3  & 45.1 
& 5521.6  & 5504.8  & 14.0 
\\ \hline
$XL_{HT}(8/6)$ max
& 2385.7 
& 2333.8 
& 62.8 
& 7706.1 
& 7684.6 
& 17.2 
\\ \hline
$XL_{HT}(8/6)$ min
& 2358.8 
& 2306.8 
& 63.0 
& 7558.8 
& 7535.5 
& 17.2 
\\ \hline
$XL_{HT}(8/8)$ max
& 2835.5  & 2741.8  & 78.2 
& 9375.0  & 9344.9  & 18.7 
\\ \hline
$XL_{HT}(8/8)$ min
& 2818.9  & 2728.8  & 77.5 
& 9329.9  & 9295.4  & 18.6 
\\ \hline
\end{tabular}
}
\caption{
Computation time (sec) of virtual machine with multiple equal processes.
}
\label{tab:multi.vht}
\EC
\renewcommand{\arraystretch}{1.0}
\vspace{-1em}
\end{table}


\begin{table}[ht!]
\renewcommand{\arraystretch}{1.3}
\BC
{
\begin{tabular}{|l|r|r|r|r|r|r|r|r|r|r|r|}
\hline
machine &  \multicolumn{3}{c|}{Math} 
      &  \multicolumn{3}{c|}{Fortran} 
\\ 
\cline{2-7}
(cores/proc.)
& real & user  & sys
& real & user  & sys
\\ \hline
$R_{HT}(8/1)$ 
& 1605.1 
& 1584.8 
& 13.9 
& 5306.9 
& 5290.8 
& 6.6 
\\ \hline
$R_{HT}(8/2)$ max 
& 1683.7 
& 1642.9 
& 15.7 
& 5345.4 
& 5329.3 
& 6.5 
\\ \hline
$R_{HT}(8/2)$ min
& 1686.9 
& 1646.2 
& 17.1 
& 5311.7 
& 5295.5 
& 6.6 
\\ \hline
$R_{HT}(8/4)$ max
& 1763.4 
& 1696.1 
& 20.6 
& 5401.6 
& 5384.9 
& 6.9 
\\ \hline
$R_{HT}(8/4)$ min
& 1765.6 
& 1700.4 
& 19.9 
& 5317.6 
& 5301.0 
& 6.9 
\\ \hline
$R_{HT}(8/6)$ max
& 2572.4 
& 2449.5 
& 29.4 
& 8063.2 
& 8039.8 
& 8.5 
\\ \hline
$R_{HT}(8/6)$ min
& 1899.1 
& 1804.9 
& 23.4 
& 7060.7 
& 7039.5 
& 8.0 
\\ \hline
$R_{HT}(8/8)$ max 
& 2882.7 
& 2707.6 
& 48.8 
& 9320.7 
& 9288.5 
& 10.0 
\\ \hline
$R_{HT}(8/8)$ min
& 2862.3 
& 2685.2 
& 47.7 
& 9291.8 
& 9261.2 
& 9.6 
\\ \hline
\end{tabular}
}
\caption{
Computation time (sec) of physical machine $R$ with HT
with multiple equal processes.
}
\label{tab:multi.r}
\EC
\renewcommand{\arraystretch}{1.0}
\end{table}


\clearpage
\newpage
\bibliographystyle{plain}


\end{document}